# Breakthrough in HAXPES Performance Combining Full-Field *k*-Imaging with Time-of-Flight Recording

K. Medjanik[1], S. V. Babenkov[1], S. Chernov[1], D. Vasilyev[1], H. J. Elmers[1], B. Schönhense[2], C. Schlueter[3], A. Gloskowskii[3], Yu. Matveyev[3], W. Drube[3] and G. Schönhense[1]

[1] *Johannes Gutenberg-Universität, Institut für Physik, D-55099 Mainz, Germany*

[2] *Imperial College London, Dept. of Bioengineering, UK*

[3] *DESY Photon Science, Notkestr. 85, 22607 Hamburg*

**Abstract**

We established a new approach to hard-X-ray photoelectron spectroscopy (HAXPES). The instrumental key feature is an increase of the *dimensionality* of the recording scheme from 2D to 3D. A high-energy momentum microscope can detect electrons with initial kinetic energies > 6 keV with high angular resolution < 0.1°. The large *k*-space acceptance of the special objective lens allows for simultaneous full-field imaging of many Brillouin zones. Combined with time-of-flight parallel energy recording, this method yields maximum parallelization of data acquisition. In a pilot experiment at the new beamline P22 at PETRA III, Hamburg, count rates of > $10^6$ counts per second in the *d*-band complex of transition metals established an unprecedented HAXPES recording speed. It was found that the concept of tomographic *k*-space mapping previously demonstrated in the soft X-ray regime works equally well in the hard X-ray range. Sharp valence band *k*-patterns of Re collected at an excitation energy of 6 keV correspond to direct transitions to the 28$^{th}$ repeated Brillouin zone. Given the high X-ray brilliance ($1.1 \times 10^{13}$ hv/s in a spot of < 20x15 μm$^2$), the 3D bulk Brillouin zone can be mapped in a few hours. X-ray photoelectron diffraction (XPD) patterns with < 0.1° resolution are recorded within minutes. Previously unobserved fine details in the diffractograms reflect the large number of scatterers, several $10^4$ to $10^6$, depending on energy. The short photoelectron wavelength (an order of magnitude smaller than the interatomic distance) "amplifies" phase differences and makes hard X-ray XPD with high resolution a very sensitive structural tool. The high count rates pave the way towards spin-resolved HAXPES using an imaging spin filter.

# 1. Introduction

Owing to the increased inelastic mean-free path, angular- or momentum-resolved photoelectron spectroscopy in the X-ray range is rapidly gaining importance for electronic structure analysis of solids. Information depths of several nm give access to the 3D electronic structure. True bulk sensitivity in the valence range has been proven using conventional spectroscopy [1] and *k*-microscopy [2], and photoemission in this regime has much potential. With increasing probing depth, the relative contribution of the surface is decreasing so that the bulk bands for samples with reactive or unprepared surfaces can be studied. Hard X-rays pave the way towards band mapping of capped surfaces, buried layers or interfaces in thin-film devices as well as in situ and operando devices. Such exciting prospects break old paradigms of photoemission and yield fascinating future perspectives.

Hard X-ray angular-resolved photoelectron spectroscopy (HARPES) faces several basic and technical obstacles. *Photoemission cross sections* drop rapidly with excitation energy above photoionization thresholds [3], which is particularly relevant for HAXPES studies of shallow core and valence levels. Alongside this drop of signal intensity, the contribution of *electron-phonon scattering* increases strongly and causes a diffuse background that eventually dominates direct interband transitions. The phonon-scattering contribution is reduced at low sample temperatures. Quantitatively, it depends on the Debye-Waller factor of the material so that criteria have been given for the feasibility of HARPES for different elements [4]. At multi keV energies, electron wavelengths are only a fraction of the interatomic distances, wherefore *X-ray photoelectron diffraction* (XPD) plays a significant role. On the one hand, XPD can yield important structural information, as discussed in comprehensive work (see, e.g., reviews [5-8]). On the other hand, valence-band photoelectron diffraction (VBPED) hampers quantitative band mapping because strong diffraction signatures can be overlaid on the band-structure patterns [9]. With increasing photon energy, the *transfer of photon momentum to the photoelectron* becomes significant. In the hard X-ray range, the resulting shift of the *k*-patterns [2] exceeds the size of a typical Brillouin zone (BZ).

The low photoemission cross section calls for optimized electron analyzers for angular-resolved HAXPES. There is a substantial difference between low- and high-energy ARPES: In the VUV spectral range the performance of an ARPES setup is limited by the maximum count rate of the electron detector. High parallelization on the detector side has led to impressive recording speeds [10]. At the same time, optimized beamline - instrument combinations offer excellent energy and angular resolution [11]. However, in the hard spectral range, the detector is no longer the bottleneck in valence band and shallow core-level photoemission, but the phase-space acceptance of the analyzer becomes the crucial criterion. Wide-angle entrance lens optics in front of hemispherical analyzers have reached large angular acceptance of up to +/-25° in the tender X-ray range [12], recorded along one line in *k*-space. A new type of entrance optics can convert the angular pattern into a *k*-pattern (of variable magnification) at the analyser entrance, thus giving access to very large angular acceptance [13].

In this paper we introduce an alternative way to significantly enhance the photoemission signal strength in the hard X-ray range. Instead of an angular-resolving detection scheme, we employ full-field imaging of the photoelectron momentum distribution in combination with time-of-flight parallel energy recording. The first pilot experiment was performed at the new beamline P22 [14] of PETRA III (DESY, Hamburg). Yielding photon fluxes up to $1.1 \times 10^{13}$ hv/s in a spot of < 20 µm dia. with a bandwidth of 450 meV (Si(111) monochromator crystal, $\Delta E/E \sim 1.3 \times 10^{-4}$) and $\sim 2 \times 10^{12}$ hv/s with a bandwidth of 155 meV at hv= 5 keV (Si(311) crystal, $\Delta E/E \sim 3 \times 10^{-5}$), this beamline provides excellent conditions for testing the new method. Higher energy resolution (down to ~50 meV) will be available in the future with the use of additional post-monochromatization.

The instrumental key feature is an increase of the *dimensionality* of the recording scheme from 2D to 3D. Common HARPES hemispherical analyzers simultaneously record a certain energy band (typically 10% of the pass energy) and the angular distribution along one line in the emission pattern ($I(E_{kin},\theta)$ recording scheme). The new technique is based on full-field imaging of the 2D ($k_x$, $k_y$) momentum distribution, as described in [15]. Instead of energy-filtering using a dispersive analyser, we implement time-of-flight (ToF) parallel detection of many energies ($I(E_{kin}, k_x, k_y)$ recording scheme). The new microscope optics was designed for initial kinetic energies up to 8 keV. The field of view in *k*-space was increased by almost one order of magnitude in comparison with the previous instruments [2,15] by a modified lens design optimized for low aberrations at high initial kinetic energies.

## 2. *k*-imaging time-of-flight HAXPES

### 2.1 Time-of-flight versus dispersive energy recording

Besides the "intrinsic" obstacles (low cross section, strong phonon scattering, diffraction), the strong retardation of the fast photoelectrons to the pass energy of the analyser constitutes a technical problem which limits performance. This can most readily be seen from Liouville's theorem, describing the conservation of the phase-space volume in particle optics. The three relevant phase-space "coordinates" are electron velocity (proportional to $\sqrt{E_{kin}}$), lateral size of the beam (quantified by the magnification M) and angular spread $\alpha$ measured with respect to the optical axis (quantified by $\sin \alpha$).

$$M \sin \alpha \sqrt{E_{kin}} = \text{const.} \qquad (1)$$

The energy resolution of a dispersive analyser is determined by the size of the entrance slit and the pass energy, cf. Fig. 1(b). Assuming a radius of the central beam of 200 mm, an energy resolution of 125 meV requires combinations of pass energy and slit width in the range of 30 eV / 3 mm to 200 eV / 0.25 mm. As a typical example, we consider a deceleration from 6.5 keV to 100 eV, the square root of the ratio yielding a factor of 8. Further, a reduction of the angular range on the sample from ~ 50° to ~15° accepted by the analyser causes another factor of 3. According to Eq. (1) the energy reduction and angular compression must be paid for by an increase of the beam size by a total factor of 24. In turn, pass energy and entrance slit of the analyser translate to maximum spatial and angular acceptance of the photoelectrons released from the sample.

Imaging ToF spectrometers (demanding pulsed excitation) do not require slits but instead require an isochrone surface with narrow width $\Delta t$ at their entrance, see Fig. 1(a). Their energy resolution is determined by the drift energy $E_d$ in the ToF section (of length *L*) and the time resolution $d\tau$ of the time-of-flight detector [16].

$$dE/d\tau = -2(2 E_d^3/m_e)^{1/2}/L \qquad (2).$$

Note that, besides *L*, this equation does not contain any geometry factors. For the instrument used in this work (*L* = 900 mm, detector 80 mm dia., $d\tau$ < 200 ps), nominal resolutions between 180 meV and 23 meV are reached at drift energies between 80 and 20 eV. The best energy resolution reached so far was 17 meV (using the identical detector in a different setup). There are no restrictions in beam size or angular spread, so long as the isochrone surfaces in the ToF section are sufficiently planar (cf. Fig. 2 for simulations of the correct lens geometry). In practice, a constant image-field curvature due to the planar detector and an additional intensity-dependent curvature due to the space-charge effect can be corrected as described in [17]. The phase-space confinement induced by the entrance slit of a hemispherical analyzer is thus replaced by the demand for sufficiently precise timing conditions of the source (for PETRA III pulse widths of ~50 ps rms) and detector (DLD resolution 150 ps).

At this point we conclude that for pulsed excitation sources, Liouville's theorem is in favour of imaging ToF instruments, owing to their linear columns lacking any slits. The principal advantage of ToF over dispersive energy recording is most apparent when aiming at very large fields-of-view at high energies (leading to huge phase-space volumes, Eq. (1)). The same argument holds when aiming at energy resolution in the few-meV range, demanding a very small entrance slit in a hemispherical analyzer. Bandwidths of < 20 meV have been reached in several soft X-ray beamlines (e.g., [18]). At beamline P22, a bandwidth of < 100 (50) meV will be possible at 3.4 (6-10) keV after installation of the channel-cut post-monochromator [14]. The ToF instrument reaches an energy resolution of ~20 meV, so the overall resolution in ToF-HAXPES experiments is limited by the photon bandwidth. The 3D recording scheme sketched in Fig. 1(a) efficiently counteracts the reduced photon flux and cross section.

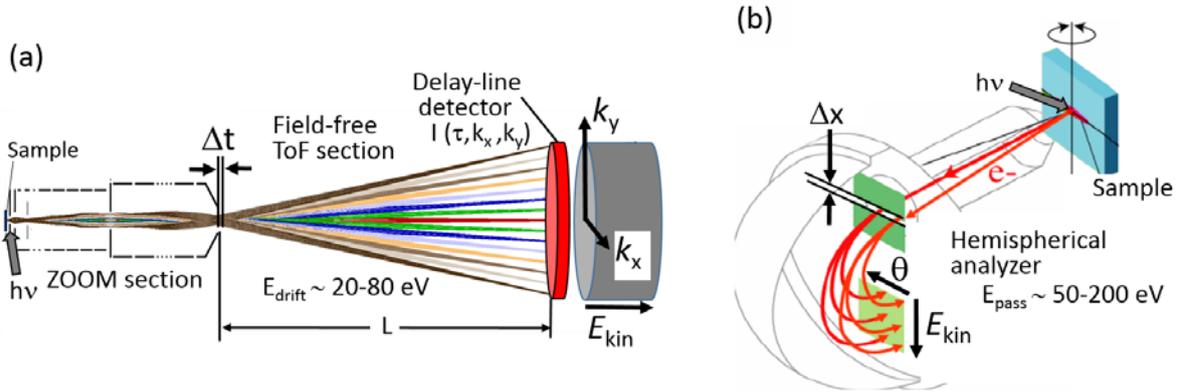

**Figure 1.** Recording schemes of a HAXPES experiment using a *k*-imaging time-of-flight spectrometer (a) in comparison with a dispersive hemispherical spectrometer (b). In (a) 3D data arrays $I(E_{kin},k_x,k_y)$ are recorded using a delay-line detector. There is no beam-confining slit (except the field aperture selecting the region of interest in real space). In (b) 2D arrays $I(E_{kin},\theta)$ are recorded and the second angular coordinate is varied either by sample rotation or by deflectors in the lens optics. The desired energy resolution demands a sufficiently "thin" isochrone surface with small width $\Delta t$ at the entrance of the ToF section (a) and a sufficiently narrow entrance slit with width $\Delta x$ at the entrance of the hemisphere (b).

*2.2 High-energy k-space microscope*

Alongside the high kinetic energy, the key feature of the new microscope is the increased diameter of the *k*-space images. A particular challenge was the development of an objective lens that keeps the image aberrations small despite very large phase-space volumes. In addition, this lens allows imaging in the conventional acceleration mode (extractor at high positive voltage), at zero extractor field and in the space-charge suppression mode using a strong decelerating field in front of the sample surface. This aspect of space-charge correction and suppression is discussed in [17,20].

*k*-space microscopy makes use of a basic concept of optics: in imaging systems the reciprocal image represents the distribution of the transversal momentum components (in mathematical language it is termed the Fourier image). Owing to $k_\parallel$-conservation in the photoemission process, the reciprocal image formed in the backfocal plane of a cathode lens directly shows the transversal momentum distribution of the valence electrons inside the crystal. A compelling advantage is that this method yields the 2D ($k_x,k_y$)-distribution in a very large region without sample rotation or scanning deflector elements. Energy recording via ToF bears the advantage that many energy surfaces are acquired simultaneously. The present measurements employed the 40-bunch mode of PETRA III with a pulse period of 192 ns. In an experiment at MAX II, we used the full 100 MHz filling pattern, i.e. 10 ns pulse period, which was sufficient for core-level XPD studies [21]. The delay-line detector (DLD) has a diameter of 80 mm and a spatial resolution of ~ 100 μm, thus resolving ~0.5 megapixels. The maximum integral count rate is 8 Mcps at a best time resolution of 150 ps [19].

Fig. 2 shows results of ray-tracing calculations for the high-energy momentum microscope. The three lens groups are schematically indicated by the ellipses, Gaussian and reciprocal planes are denoted. A special objective lens designed for minimum spherical aberration forms the first *k*-image in its back focal plane (BFP). The electron beam is transformed to the scattering energy of the spin filter (not shown here) by zoom optics 1 and to the desired drift energy in the imaging ToF spectrometer by zoom optics 2. Photoelectron momentum maps are taken simultaneously in an energy interval of several eV width, limited by the chromatic aberration of the instrument. The simulations were based on the small footprint (~20 µm) of the photon beam at P22, yielding a large depth of focus with > 6 eV usable energy range. The large field of view in reciprocal space is imaged practically without aberrations. Time markers with 10 ns spacing (dots) reveal planar isochrones, an essential precondition for good energy resolution and no "crosstalk" between longitudinal and transversal momentum components.

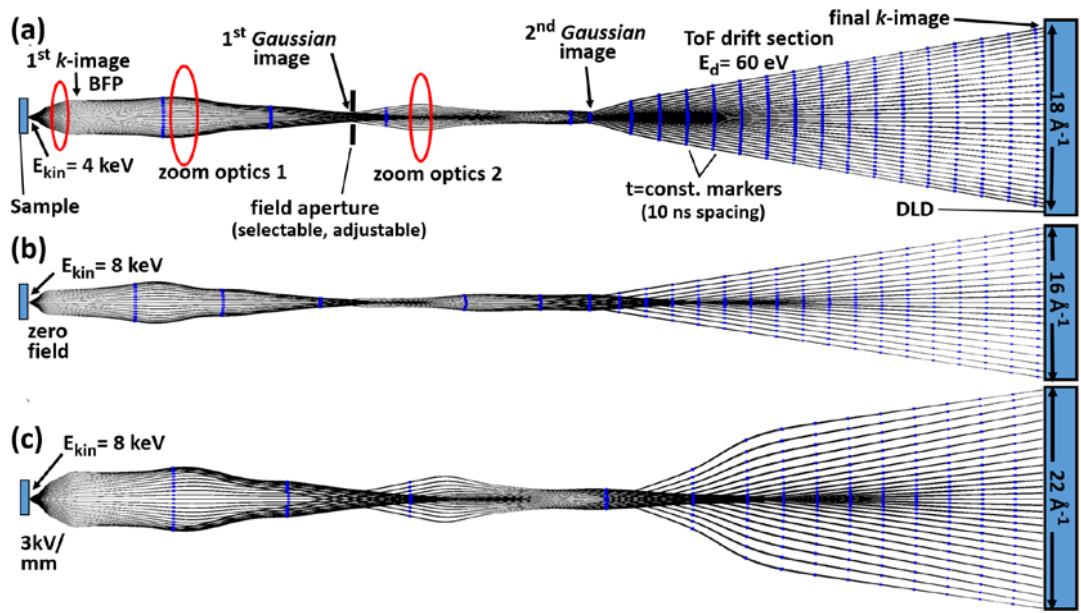

**Figure 2.** Simulated electron trajectories for the high-energy optics. Large momentum ranges with diameters of 18 Å$^{-1}$ (a), 16 Å$^{-1}$ (b) and 22 Å$^{-1}$ (c) are imaged to a delay-line detector (DLD) of 80 mm diameter. Three lens groups focus and decelerate the beam; an adjustable field aperture serves for selection of the region-of-interest. Initial kinetic energies are 4 keV (a) and 8 keV (b,c); drift energy is 60 eV for all. Electric field at the sample surface was 3 kV/mm (a,c) and zero (b). Blue dots are time markers (10 ns spacing).

The example of Fig. 2(a) was calculated for an initial kinetic energy of 4 keV and an acceptance of 18 Å$^{-1}$ diameter in *k*-space. The new objective lens allows working without an acceleration field between sample and anode; in this mode the lens elements behind the anode focus the beam. An example of this mode is shown in Fig. 2(b) for an initial energy of 8 keV. Despite zero extractor field, *k*-fields as large as 16 Å$^{-1}$ can be recorded without visible aberrations. For the same kinetic energy at the sample, a moderate extractor field of 3 kV/mm yields *k*-fields up to 22 Å$^{-1}$ on the 80 mm image detector (Fig. 2(c)), and even larger fields for larger detectors. For the given instrument, we gain an order of magnitude in total intensity in comparison with measurements using the low-energy ToF *k*-microscope in the soft X-ray range [2].

Key features for electronic band mapping are the increased total intensity and the larger information content by imaging many BZs simultaneously. An important advantage of the modified optics concerns the rapid recording of X-ray photoelectron diffraction (XPD) patterns. Rich diffraction features appear beyond typically 3 Å$^{-1}$ radius (cf. examples in Fig. 6). This outer region is principally inaccessible using existing low-energy *k*-microscopes.

*2.3 Time-of-flight recording in the X-ray range*

Since ToF instruments are high-pass filters, one might suspect that they are not suitable for core-level studies in general due to overlap of the signal with other core levels or the valence electrons or contributions from higher-order radiation from the beamline. In this section, we discuss the special conditions under which such undesired signals appear and possible solutions for their suppression. For the given conditions ($L$ = 900 mm, d$\tau$ < 200 ps), the energy dispersion according to Eq. (2) is $\Delta E$/meV = 0.255 ($E_d$/eV)$^{3/2}$. Owing to the small photon footprint (Fig. 3(a)), the measurements shown in Section 3 were performed with open field aperture. The DLD records all counting events in the selected $E$-$k_\parallel$ region confined by the constant energy surface at the Fermi energy $E_F$ and the diameter of the k-region of interest. The binding energy (with respect to $E_F$) is determined as in classical photoemission ($E_B$ = h$\nu$ - $E_{kin}$ - $\Phi$; $\Phi$ work function) utilizing the relation $E_{kin}$= ½ $m_e$(L/$\tau$)$^2$ (L, $\tau$ and $m_e$ being the length of the drift section, the time of flight and the electron mass, respectively). For absolute ToF measurements, $\tau$ can be referenced to the photon signal itself (see Fig. 3(b,c)). In practical work, the energy scale is calibrated by two measurements with slightly different sample bias yielding dE/d$\tau$ and observation of the Fermi edge yielding the zero reference for $E_B$. The delay-line detector records the three coordinates (x-position, y-position, and arrival time $\tau$) of each counting event; the events are accumulated in the ($E_B$,$k_x$,$k_y$)-voxels of a 3D data array. For visualization, either the full 3D object is displayed (cf. [22]), or sections can be cut along any plane (as in Figs. 3(d,f,g) and 4(b-g)).

The kinetic energy in the drift section $E_d$ is given by

$$E_d = h\nu - \Phi_{sa} - E_B - eU_{sa} + eU_d + \Delta\Phi \qquad (3)$$
$$= eU_d + \Delta\Phi \quad \text{for the condition} \quad U_{sa} = (h\nu - \Phi_{sa} - E_B)/e = E_{kin}^{initial}/e$$

Where $\Phi_{sa}$ is the work function of the sample and $\Delta\Phi$ the work function difference between sample and drift tube; $U_{sa}$ and $U_d$ are the voltages at the sample and the drift tube. Then, only electrons in a band between $E_F$ and $E_B$ = e$U_d$ are transmitted, with the topmost 6 eV being well focussed. By using another lens as high-pass cut-off it is possible to restrict the energy interval further, as described in [23]. A ToF instrument also records photoelectrons from higher orders (2h$\nu$, 3h$\nu$) of the monochromator / undulator. Photoelectrons released by higher-order photons are much faster and thus have a shorter time of flight $\tau$. If $\tau$ is shorter by one period of the exciting photon pulses (here 192 ns in the 40-bunch mode of PETRA III), they appear in the same time interval as the photoelectrons from first order. A typical example is displayed in Fig. 3(f-h), showing the coincidence of the valence-band pattern of Mo (excited by h$\nu$= 3100 eV) with the Mo 2s and 2p$_{1/2,3/2}$ core-level signals (binding energies 2866, 2625 and 2520 eV, respectively) released by a higher-order admixture. The signal of this admixture incidentally falls into the same time interval.

The easiest way to shift a higher-order contribution out of the spectrum is to change the photon energy, since the time of flight of the true and higher-order signals respond differently. A change of the drift energy also leads to a separation from such unwanted signals. Being much faster, the higher-order electrons are essentially unfocussed and pass the microscope column as a pencil beam, confined by the apertures in the ray path (white area in (f)). The different focussing conditions offer an alternative way to use the apertures as a beam stop for the fast electrons, such that electrons with the proper energy can pass the apertures. Since the higher-order signals are not focussed, they appear unstructured and can be eliminated numerically if their intensity is not too high. In the future, an additional dispersive element (as a coarse bandpass filter) will eliminate these contributions completely. Fig. 3(i) shows valence-band spectra of NbSe$_2$, taken with two different monochromator crystals. For Si(111) and Si(311) we determine bandwidths of the X-ray beam of 450 and 155 meV, respectively (cut-off profile at $E_F$; 16-84% criterion).

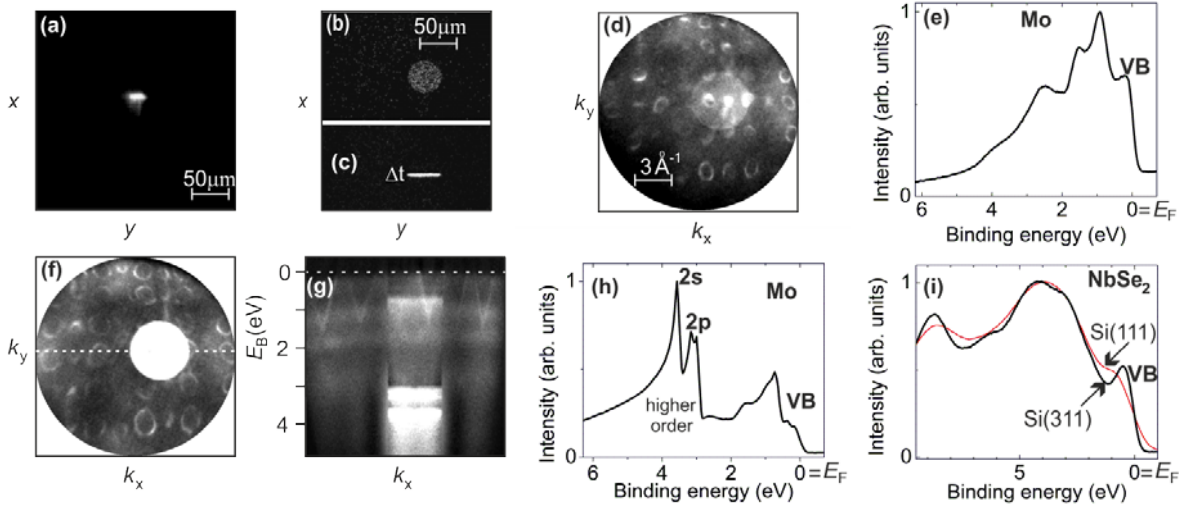

**Figure 3.** (a) Footprint of the photon beam, imaged using photoelectrons emitted close to the Fermi edge. (b) Signal of photons scattered at a surface defect; (c) corresponding time spectrum. (d) Momentum distribution of valence-band photoelectrons from Mo(110) at hν= 3100 eV (cut corresponds to the Fermi energy) and (e) corresponding binding-energy spectrum. (f) Cut of the same data set at a binding energy of 3.2 eV, where the 2s / 2p core-level signals (bright circle) originating from a third-order admixture (hν= 9300 eV) coincides with the valence-band pattern (details, see text). (g) Corresponding $E_B$-vs-$k_x$ section and (h) corresponding ToF-spectrum (binding energy scale corresponds to the valence band VB). (i) Valence-band spectrum of $NbSe_2$ at hν= 5320 eV, taken with the Si(111) and Si(311) monochromator crystals; thin and full spectrum, respectively.

## 3. Applications
### 3.1 Mapping of bulk valence-bands

One of the key questions addressed in the first pilot experiment was whether mapping of the electronic structure using *k*-microscopy works in the hard X-ray range. Although valence-band features had been observed at high X-ray energies [1,4,24,25], it seemed that they are strongly masked by quasi-elastic phonon scattering. The diffuse background of such electrons exhibits a characteristic spectral distribution given by the "matrix-element weighted density of states" [26-28]. Moreover, it was not clear whether the $k_z$-resolution stays sufficiently high in the several-keV range. This is a necessary precondition for precise mapping of bulk bands. The transfer of photon momentum $k_{hv}$ to the photoelectron leads to a substantial shift of the *k*-pattern. The interplay of this shift with X–ray diffraction of the photoelectrons from the valence-band is still in an early stage of understanding [9]. Last but not least, typical count rates of conventional HAXPES spectrometers seemed prohibitively low for extended 4D mappings. This would require 3D data arrays with good counting statistics taken at many different photon energies in order to vary $k_z$. In this Section we will show that mapping of the 4D spectral function ρ($E_B$,***k***) in a tomography-like manner as introduced in the soft X-ray range [2] still works for hard X-rays, if the momentum conservation including the photon momentum is properly accounted for.

Fig. 4(a) shows a quantitative scheme of direct transitions in Re at photon energies between 2600 and 6000 eV, leading to the region between the 18$^{th}$ and 28$^{th}$ repeated BZs along $k_z$. The plot is to scale with $k_z$ and $k_y$ being quantified by the reciprocal lattice vectors along the directions Γ*A* and Γ*M* for the hcp metal Re, ***G*$_{0001}$** (1.410 Å$^{-1}$) and ***G*$_{11-20}$** (2.276 Å$^{-1}$), respectively. The dashed rectangle in the lower inset denotes the cut of the first BZ, shown in (h). Since the $k_z$-axis points perpendicular to the surface, the background pattern in (a) is not obtained from a single measurement but from a series of 3D arrays recorded at more than 20 different photon energies corresponding to different $k_z$ values (see below). The transfer of photon momentum $k_{hv}$ to the photoelectron causes a strong displacement of the center of the final-state energy isospheres from the origin **k**=(0,0,0) (marked by **+**) by as much as 3.04 Å$^{-1}$ at 6 keV. Due to the impact angle of 22° from the surface, the main shift ($|k_{hv}|\cos22°$= 1.24 $G_{11-20}$) acts in the lateral direction $k_y$. $|k_{hv}|\sin22°$= 0.81 $G_{0001}$ acts in the perpendicular direction. $k_{hv}$ is proportional to hν, whereas the photoelectron momentum $k_f$ increases with the root of the energy (the equation denotes the final-state momentum $k_f$ inside of the solid):

$$k_{hv}= 2\pi\nu/c, \qquad k_f = (1/\hbar)\sqrt{2m_{eff}E_{final}}, \qquad \text{with} \qquad E_{final} = h\nu - E_B + V_0^* \qquad (4),$$

where binding energy $E_B$ and inner potential $V_0^*$ are referenced to the Fermi energy. In the hard X-ray range, the final-state effective mass $m_{eff}$ is identical to the free electron mass [1]. Growing linearly with hν, the photon-momentum induced shift becomes very substantial at high energies. For Re at 6 keV, it shifts the center of the final-state sphere transversally by more than a full BZ. The graphical model in Fig. 4(a) reveals the resulting dissymmetry in the observed photoemission patterns (top insets). In *k*-microscope images, this rigid shift by $k_{hv}$ is directly observable as discussed in [2].

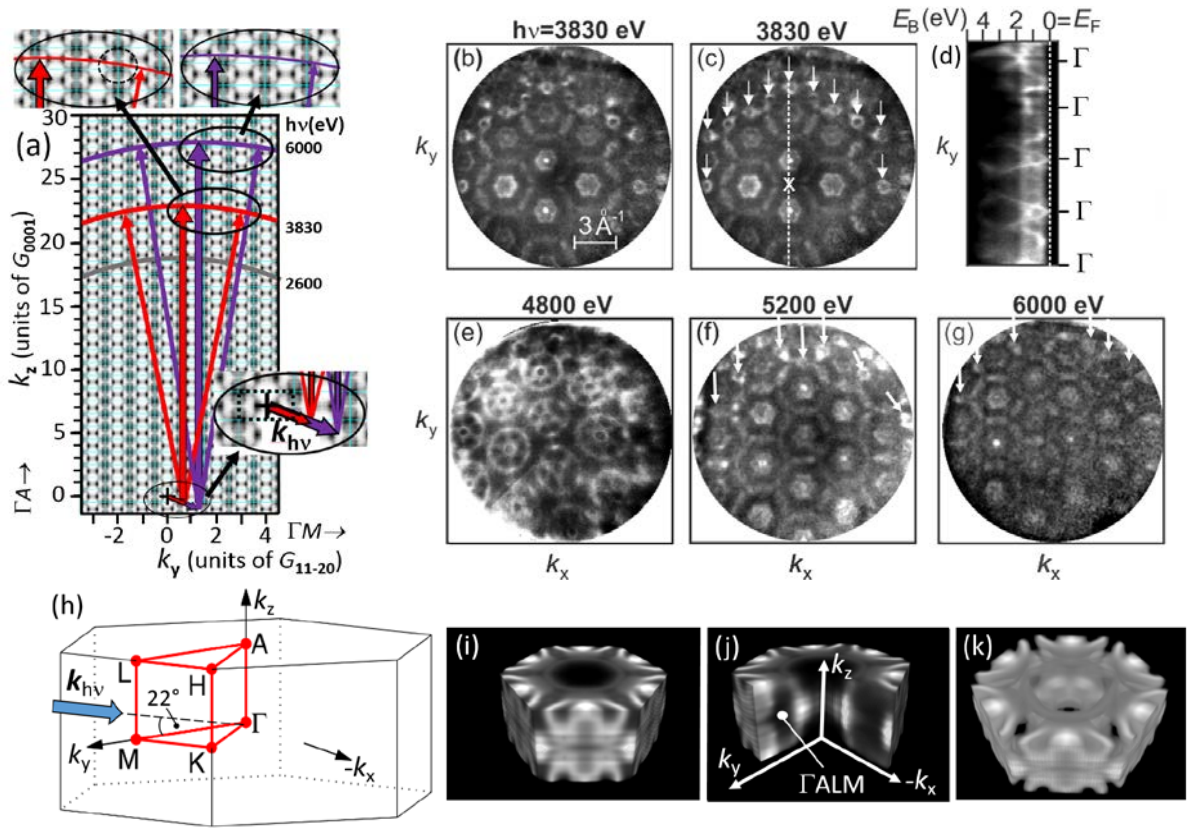

**Figure 4.** (a) *k*-space scheme of direct transitions from the valence-band of rhenium at hν= 2600, 3830 and 6000 eV to free-electron-like final states. The background pattern in (a) is a measured Fermi-surface cut in this plane, the pattern is to scale with $k_z$ and $k_y$ quantified in multiples of the reciprocal lattice vectors $G_{0001}$ and $G_{11-20}$ for Re. The dotted rectangle in the lower inset marks the first BZ. The three transitions lead to final-state isoenergetic spheres crossing the 18$^{th}$, 23$^{rd}$ and 28$^{th}$ repeated BZ along $k_z$. $k_{hv}$ is the photon momentum which lies in the drawing plane and leads to a displacement of the centers of the final-state spheres (see lower inset). (b) Measured large-area *k*-space map for the *d*-bands of Re taken at hν= 3830 eV (T ~20 K) showing a cut through the center, first ring and second ring of repeated BZs. (c) The same with marked features, see text. (d) $E_B$-vs-$k_y$ cut along the dashed line in (c) showing the band dispersion. (e-g) *k*-space maps at higher photon energies as given on top of the panels. (h) Brillouin zone with high-symmetry points and geometry of photon impact. (i-k) Measured Fermi surface in different perspectives, (j) denotes the cut leading to the background pattern in (a).

Fig. 4(b) shows a $k_x$-$k_y$ cut (at $E_F$) through the measured data array at hν= 3830 eV; all patterns shown were measured at low sample temperature (T ~20 K). Each of the sixfold rings represents one Brillouin zone (BZ); we see one in the center, a first ring with six and a second ring with 12 adjacent BZs, making up 19 BZs observed in parallel. Accumulation time was ~15 min. at a total count rate of ~$10^6$ counts/s. A slight image-field curvature (due to the large field of view) and Lorentzian deformation due to a weak but significant space-charge shift was corrected using the algorithms described in [17]. The diffuse background due to phonon scattering was removed by the procedure described in [1], as well as an additional effect due to valence-band photoelectron diffraction as discussed in [9]. Details on these procedures can be found in the given references.

It is eye-catching that in each of the images, different BZs look differently. Moreover, the patterns do not exhibit the full sixfold symmetry which we would expect for Re(0001). Instead, a two-fold symmetry modulates the pattern of sixfold rings. The center of this apparent twofold symmetry is marked by a cross in Fig. 4(c). The reason for this - at first sight puzzling - deviation from the expected symmetry becomes clear when we take into consideration that the final states lie on a curved energy isosphere, see details on top of (a). Due to the increased field of view of the present instrument, the curvature of the intersection contour of the sphere and the periodic $k$-space pattern becomes substantial. In the previous experiments in the soft X-ray range, focussing on roughly one BZ, this effect was just a small correction. The "symmetry point" marked by X in Fig. 4(c) corresponds to the top of the sphere and with increasing distance from this point, the $k_z$-value is reduced. In turn, the cut runs through different features of the 3D Fermi surface. As a guide to the eye, we marked the small circles in the Fermi-surface cut with arrows in (c). These circles appear when the sphere intersects a small Fermi-surface pocket located about midway between the ΓMK and ALH symmetry planes (cf. (h)). In the observed patterns, these circles lie on a ring because these points have the same distance from the top point X, see dashed circle in the upper left inset of (a).

Note that the top of the sphere is not given by a Γ-point but is defined by the size and direction of the photon momentum vector. The role of $k_{hν}$ in HARPES is not at all trivial and influences the observed patterns substantially. The final-state sphere is just rigidly shifted in 3D $k$-space. However, concerning the observed pattern, this shift shows up in a complex way which can only be understood in a full 3D $k$-space transition model. Fig. 4(a) shows the plane which contains $k_{hν}$, hence perpendicular to this plane there is no photon momentum component. This causes the apparent 2-fold symmetry. Experimentally, the direction perpendicular to the surface can be identified by the profile of the space-charge induced Lorentzian energy shift [17]. The Γ-point at $k_∥$ = 0 is displaced from the perpendicular direction by $|k_{hν}|$cos22° as discussed above.

The curvature of the final-state sphere is also visible in the $E_B$-vs-$k_y$ cut (d). The centers of the different BZs are denoted by Γ (keeping in mind that this is not exactly the bulk Γ-point). We see that the downward-dispersing band only reaches the Fermi energy in the two points next to "X" . The 4D nature of the spectral density function ρ($E_B$,$k$) along with the a-priori curved final-state isosurface and the broken symmetry by the action of the photon momentum lead to a complex behaviour. (e-g) show maps measured at higher photon energies. (e) corresponds to a binding energy of 2 eV, whereas all other panels in Fig. 4 show cuts at the Fermi energy. For (f) and (g) we have chosen energies that lead to equivalent "planes" in different repeated BZs. Since the curvature of the sphere is reduced with increasing hν, the small circles are shifted outwards (white arrows in (c,f,g)), thus confirming the model proposed in (a). Due to the curvature of the final-state sphere, the large field of view crossing several BZs at different levels allows, in principle, mapping of the spectral function in measurements at few photon energies or even just one photon energy (if the size of the BZ along $k_z$ is small).

Figs. 4(i-k) show the measured Fermi surface of Re, i.e. the spectral-density function at the Fermi energy (array ρ($E_F$,**k**)). First, individual energy isosurfaces at $E_F$ have been obtained from measurements at ~20 different photon energies, crossing a full BZ. Then, these arrays are concatenated to form the complete 3D Fermi surface. These measurements have been made in the soft X-ray range (at PETRA-beamline P04), with its intrinsically higher energy resolution. This experimental Fermi surface was the basis for the background pattern in Fig. 4(a) which is a cut at $k_x$ = 0 (i.e. the ΓALM-plane) of the periodically-repeated Fermi surface. Note that in (a) the contrast is inverted for clarity (dark denotes high spectral density).

### 3.2 Core-level spectromicroscopy

The core-level signals carry a similarly rich information content as the valence-band patterns. In particular, their lateral intensity distribution is recorded in extremely short time: in the present experiments, a few minutes for a pattern with good statistics. The lateral distribution represents the photoelectron diffraction pattern which yields valuable structural information. The core-level spectra give chemical information, both complementing the electronic information of the valence-band data. Exploiting the real-space imaging capability, microspectroscopy is possible with spatial resolution better than the photon footprint by using size-selectable *field apertures* in an intermediate real image plane [22]. Regions-of-interest down to the micrometer range can be selected. However, this high spatial resolution comes at the expense of a smaller diameter of the *k*-pattern, as discussed by Tusche et al. (cf. [15], figure 3).

Fig. 5 shows a collection of core-level spectra along with the corresponding angular patterns (more precisely, *k*-distributions). The change from valence-band to a core level only requires setting the sample bias $U_{sa}$ to the proper value, according to Eq. (3). For very large changes of the kinetic energy, a slight refocusing of the objective lens might be necessary. For the Re sample, we recorded spectra and *k*-patterns of the 4f doublet ((a,b), $E_B$=40.5/42.9 eV), the 5s line ((c,d), 83 eV), the $4d_{5/2}$ line ((e,f), 261 eV), the $4p_{3/2}$ line ((g,h), 447 eV) and the $3d_{5/2}$ line ((i,j), 1883 eV). Except for the rather weak 5s line, all core level spectra show excellent signal-to-background ratios. The intensity scales show the true zero, i.e. despite the high-pass characteristic the background levels are not larger than in a hemispherical analyser. The spin-orbit splitting of Re 4d (13 eV), 4p (72 eV) and 3d (66 eV) is too large to fit into the recorded energy interval at the given settings. For these three cases, only the line with the higher total angular momentum is recorded, along with all *k*-patterns correspond to this line. Multiplets with large splitting could be recorded at higher drift energies. The *k*-distributions are very pronounced and rich in details. They arise due to photoelectron diffraction, as will be discussed in the next Section.

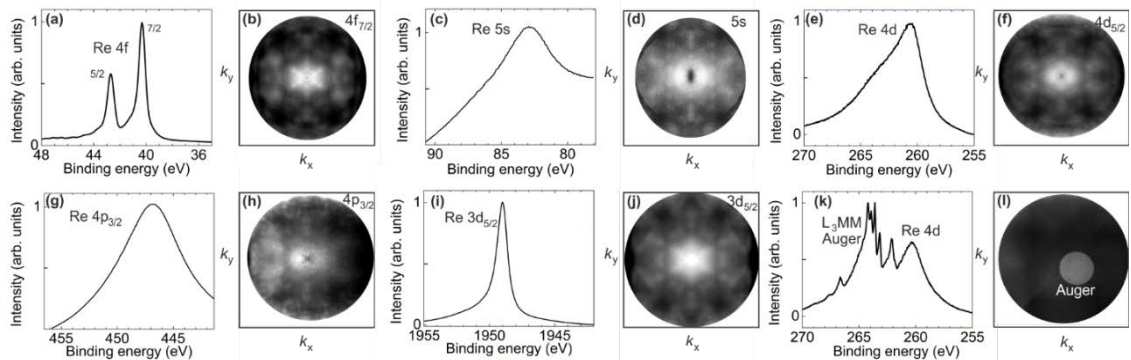

**Figure 5.** (a-j) Core-level spectra and corresponding *k*-distributions of Re 4f, 5s, 4d, $4p_{3/2}$ and $3d_{5/2}$. (k,l) Group of $L_3$MM Auger lines originating from 3rd order radiation, falling onto the wing of the Re 4d signal. Note that the binding energy scale in (k) refers only to the 4d signal (hν= 3750 eV); the Auger lines originate from 3rd order radiation (hν= 11250 eV). Except for the weak Re 5s signal, all core lines show a good signal-to-background ratio and the *k*-distributions exhibit pronounced structures.

The fortuitous coincidence of the investigated spectrum with a signal originating from higher orders of the monochromator Fig. 3(e-h) can also happen in core-level spectra. An example is shown in Fig. 5(k), where a group of L$_3$MM Auger lines falls onto the low-energy wing of the Re 4d line. The higher energy of the Auger signal is evident from the small circular region in which this signal is present (the dashed spectrum was taken in the area of the bright circle in (i)). It is straightforward to determine the energy of this Auger signal. At the given photon energy of 3750 eV, the kinetic energy of the 4d$_{5/2}$ electrons is 3484 eV (assuming a work function of 5 eV). These electrons are then retarded to 60 eV drift energy in the ToF section (i.e. their energy is reduced by 3424 eV). The Auger electrons arrived one period of the photon pulses, i.e. 192 ns, earlier. The fast electrons from photon pulse #n coincide in time with the valence electrons from photon pulse #n-1. Using the simulation program for the proper lens settings, we arrive at an initial kinetic energy of the Auger electrons of ~6000 eV, following photoemission from an inner shell by 3$^{rd}$ order photons with hν= 11250 eV. Consulting Auger tables we arrive at the L$_3$MM cascade with kinetic energies of ~ 5970 eV (~6219 eV) for the L$_3$M$_2$M$_5$ (L$_3$M$_3$M$_4$) transition. Note that the energy scale of the Auger signal is strongly stretched due to the much higher drift energies of these electrons.

### 3.3 X-ray Photoelectron Diffraction (XPD)

The full-field acquisition scheme of a time-of-flight *k*-microscope is ideal for rapid recording of XPD-patterns without changing the sample geometry. Core-level signals are selected by setting the proper sample bias, such that the binding-energy region of interest is accumulated (cf. Eq. (3)). In other words, the kinetic energy $E_{kin}$ of the core-level signal is shifted exactly to the point of best focusing of the electron optics. As discussed in Section 2.3, faster electrons originating from higher-lying core levels, from the valence band or from admixture of higher orders in the photon beam usually appear in different time slices. Hence they can be shifted out of the spectrum. Electrons slower than $E_{kin}$-eU$_d$ are cut off by the high-pass action of the drift tube.

A sequence of photoelectron diffraction patterns for the Re 4$f_{7/2}$ core-level signal taken between hν= 2800 and 6000 eV is shown in Fig. 6. These small-angle diffractograms show a surprisingly rich structure, especially at high energies. Moreover, this structure varies strongly with photon energy. For these measurements, attenuators had to be inserted in the photon beam reducing the count rate to ~2 Mcps to avoid multi-hit events. Such events cannot be properly analysed by the DLD and thus increase the diffuse background (the photon pulse rate was 5 MHz). At these conditions, diagrams like those in Fig. 6 were obtained in ~10 minutes. Using denser filling patterns of the storage ring, this problem of the Poisson statistics would be eliminated and the 8 MHz count-rate capability of the detector could be fully exploited. In the future, multi-line DLDs [19] with highly-parallelized recording architecture (and high multi-hit capability) will allow for much higher total count rates. The space charge problem is less serious when recording XPD because "true" XPD-patterns can be extracted much easier than valence-band patterns, even when Coulomb repulsion in the beam leads to a strong Lorentzian deformation [17] of the raw data.

In the series shown in Fig. 6, the kinetic energy varies by 3200 eV. Hence there is a slight systematic change in the field-of-view due to the chromatic aberration of the objective lens. The energy dependence of this effect is known and the images have been calibrated, see scale bars in (a) and (f). In order to observe a larger radius, the symmetry center of the patterns was shifted off the image center, and the measured distributions were symmetrized. The electric vector of the incoming X-ray beam is oriented 22° off the surface normal. The as-measured data show that this symmetry breaking only has a minor effect on the observed patterns. The diffractograms reflect the symmetry of the Re crystal. Translated into real-space coordinates, the diameters of the *k*-images correspond to angular ranges varying between 0-13° at 2800 eV and 0-11° at 6000 eV. Some of the fine features have widths < 0.1 Å$^{-1}$, corresponding to < 0.1°.

XPD data in literature have been obtained using conventional XPS analyzers scanning large polar angular ranges of typically 0-60°. 2D patterns are obtained by azimuthal rotation of the sample. The photon-energy dependence is often only studied in normal emission (for recent reviews, see [5-8]). Under these conditions, XPD patterns are dominated by Kikuchi lines and strong forward scattering along atom rows. In the small angular range of our experiment, forward scattering from atom rows can be excluded (except the row perpendicular to the surface). The pattern at hν= 3400 eV shows a significant intensity enhancement in the center that might indicate constructive forward scattering at this photoelectron wavelength. The 6000 eV pattern shows a system of dark straight lines with sixfold symmetry that might be interpreted as Kikuchi lines. However, no pronounced straight lines appear in all other patterns. Most features, e.g. the leaf-shaped dark regions in (a), the filigree-like central regions in (d-f) with small spots and circles must be explained by a different mechanism. This rich structure is a new kind of fingerprint of multiple scattering of the outgoing photoelectron wave.

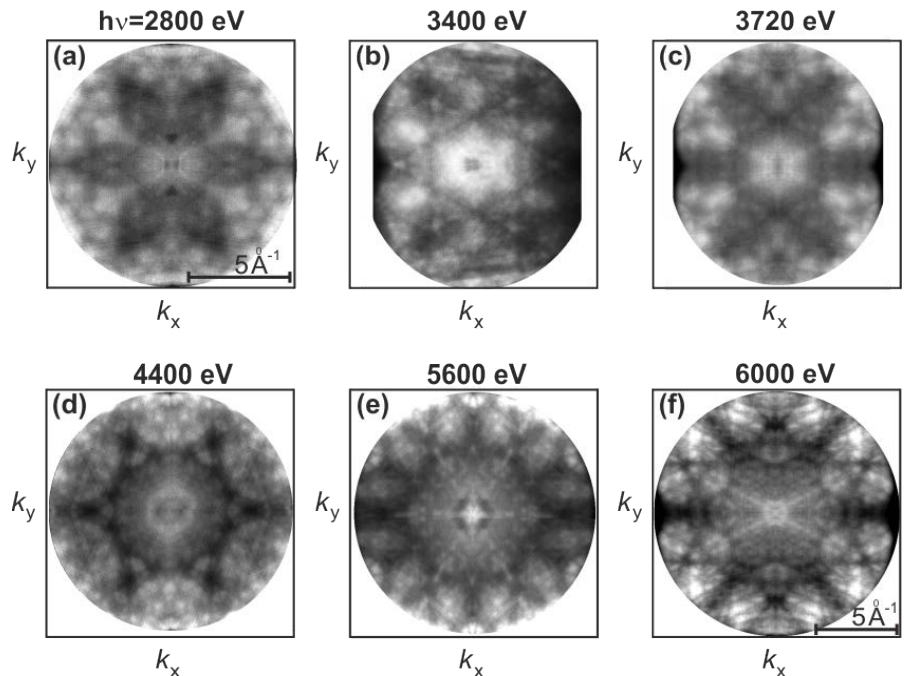

**Figure 6.** Sequence of photoelectron diffraction patterns of the rhenium $4f_{7/2}$ core-level signal at selected photon energies between 2800 and 6000 eV. The *k*-scale varies with energy, see scale bars in (a) and (f). The corresponding range of polar angles varies between 0-13° in (a) and 0-11° in (f). The contrast has been increased in order to emphasize fine features.

It is eye-catching that the sharpness of the features increases in the sequence (a-f). With increasing photon energy two effects are important: the photoelectron wavelength is reduced from 23.17 pm (in a) to 15.83 pm (in f) and the inelastic mean free path $\lambda_{IMFP}$ increases from ~3 nm (in a) to ~5.4 nm (in f). The spherical wave excited from the isotropic (or statistically isotropic) initial state interacts with all scatterers within an effective volume confined by inelastic scattering. All atoms within this volume emit spherical waves with amplitudes weighted by their distance from the emitter atom (at a given energy all have identical scattering phases). Significant diffraction contributions can be expected until the amplitudes of the scattered wave have dropped to 10% of the initial intensity, i.e. for a linear distance from the emitter atom up to 2.3 $\lambda_{IMFP}$. Given a nearest-neighbour distance of the Re atoms of 0.276 nm, we estimate that about 70,000 (in a) to 400,000 atoms (in f) contribute to the observed diffraction pattern. The coherent superposition of that many spherical waves, all contributing weighted by their distance from the emitter atom, can be expected to lead to such rich structures like those in Fig. 6. The strong increase in the number of atoms explains why the small features get sharper at higher energies, cf. sequence (d,e,f).

Since the photoelectron wavelength is more than an order of magnitude smaller than the nearest-neighbour distance, small changes in the electron wavelength or the sample structure lead to substantial differences in the diffraction pattern, because the corresponding phase difference is "amplified" by this factor. The simplest interference condition is constructive forward scattering normal to the surface. At 3400 eV, the final-state momentum along $k_z$, corrected by the photon momentum according to Eq. (4), is $k_f$ = 20.5 $G_{0001}$. This condition corresponds to the Bragg condition for constructive interference in normal emission (for details, see eq. (6) in [9]). This explains the enhanced intensity in the center of (b). A further quantitative analysis of these patterns requires more elaborate computational analysis and is beyond the scope of the present paper.

## 4. Summary and Conclusions

This paper introduces a highly effective way to cope with the weak signals in hard X-ray angular-resolved photoelectron spectroscopy (HARPES). Instead of the conventional angular-resolving $I(E_{kin},\theta)$ detection scheme, we employ full-field $I(k_x,k_y)$ imaging and time-of-flight ($\tau$) parallel energy recording. This 3D recording scheme effectively counteracts the dramatic drop in photoemission cross sections and increase of quasi-elastic but $k$-randomizing electron-phonon scattering cross sections. A special objective lens and retarding zoom optics have been developed, allowing for very large $k$-fields of view, an order of magnitude larger area than recorded by the existing low-energy instruments. The main goal of the study was to investigate the capability for valence-band mapping, and benchmark the results against the pioneering experiment of Gray et al. [1] at h$\nu$= 6 keV. Photoelectrons from the valence band are the fastest and thus easily accessible by a ToF spectrometer, which is inherently a high-pass filter. A crucial question prior to the experiment was whether deeper-lying core-levels are accessible as well. This would pave the way towards fast recording of X-ray photoelectron diffraction (XPD) patterns, simultaneously with valence-band momentum mapping.

The performance of instrument and method was elucidated in a first pilot experiment at the new dedicated HAXPES beamline P22 at PETRA III (DESY, Hamburg) [14]. We recorded $I(E_B,k)$ data arrays at many photon energies in the range of 2.6 to 6 keV. Valence-band $k$-maps of excellent quality comprising 19 Brillouin zones have been recorded at low sample temperatures. Thanks to the high brilliance of beamline P22 (1.1x10$^{13}$ h$\nu$/s in a spot of ~20 $\mu$m), valence-band signals reached up to several million counts per second in the $d$-band complex of transition metals (Re, Mo). At such conditions, 3D bulk BZs can be mapped in a tomography-like manner in a few hours by varying the photon energy and exploiting direct transitions into free-electron-like final states.

Core-level spectroscopy is easily possible as demonstrated for several inner shells. Signals from faster electrons (originating from higher-order radiation of the beamline) can be identified by their different focusing properties. We discussed strategies to eliminate such admixtures in the true signal. Full-field core-level XPD patterns are recorded within minutes. Their richness in fine details reveals a new appearance of XPD at very high angular resolution of < 0.1° which is neither due to Kikuchi lines or forward scattering along off-normal directions (being the origin of the most prominent features in conventional XPD at larger angles), nor due to transfer of reciprocal lattice vectors as discussed in [9]. XPD originates from the far-field interference of many spherical waves emerging from all atoms located within a scattering volume with radius ~2.3 $\lambda_{IMFP}$ around the emitter atom ($\lambda_{IMFP}$, inelastic mean free path). The fine details in the observed patterns reflect the large number of scatterers, several 10$^4$ to 10$^6$, depending on energy. The short photoelectron wavelength (more than an order of magnitude smaller than the nearest-neighbour distance) "amplifies" the sensitivity of the interference pattern and is the key factor for the richness in details. In turn, the patterns react strongly on small changes of photon energy and can be expected to be a sensitive probe of structural changes.

On the technical side, the new objective lens works with zero field and thus opens the path to samples with corrugated surfaces, small crystallites and 3D micro- /nanostructures. The retarding mode provides an efficient suppression of space-charge interaction with the large amount of secondaries in the photoelectron distribution at high photon energies. Due to the curvature of the final-state sphere, the large field of view with many Brillouin zones allows, in principle, mapping of the spectral function in a measurement at a single photon energy.

Full-field *k*-imaging combined with ToF recording established an unprecedented data-acquisition speed in hard X-ray photoemission. The high-energy *k*-microscope detects initial kinetic energies in the HAXPES range up to ~ 8 keV. A new type of objective lens, optimized for low spherical aberrations at high kinetic energies, can image areas in *k*-space about an order of magnitude larger than all previous designs. This lens can work at zero extractor field, thus eliminating one of the serious obstacles of cathode-lens instruments for corrugated surfaces (as might result from imperfect cleaving or 3D nanoobjects). Moreover, it can even operate in an "inverted" mode, using strongly-retarding fields in front of the surface in order to eliminate the deterministic part of space-charge interaction, as will be discussed in [20]. Last but not least, the high count rates pave the way towards spin-resolved HAXPES using an imaging spin filter.

### Acknowledgements

Our thanks are due to A. Oelsner and M. Ellguth (Surface Concept GmbH, Mainz, Germany) and C. Tusche (FZ Jülich, Germany) for continuous support of this development and to K. Ederer (DESY) for valuable technical support. We thank K. Rossnagel for fruitful discussions and constructive comments. Funding by BMBF (05K16UM1, 05K16UMC) and DFG (Transregio SFBs TR49 and TR173) is gratefully acknowledged.

### References

[1] A. X. Gray, C. Papp, S. Ueda, B. Balke, Y. Yamashita, L. Plucinski, J. Minár, J. Braun, E. R. Ylvisaker, C.M. Schneider, W. E. Pickett, H. Ebert, K. Kobayashi and C. S. Fadley, *Probing bulk electronic structure with hard X-ray angle-resolved photoemission,* Nature Materials **10**, 759 (2011).
[2] K. Medjanik, O. Fedchenko, S. Chernov, D. Kutnyakhov, M. Ellguth, A. Oelsner, B. Schönhense, T. R. F. Peixoto, P. Lutz, C.-H. Min, F. Reinert, S. Däster, Y. Acremann, J. Viefhaus, W. Wurth, H. J. Elmers and G. Schönhense, *Direct 3D mapping of the Fermi surface and Fermi velocity,* Nature Materials **16**, 615 (2017).
[3] M.B. Trzhaskovskaya, V.G. Yarzhemsky, *Dirac–Fock photoionization parameters for HAXPES applications,* Atomic Data and Nuclear Data Tables **119**, 99–174 (2018).
[4] C. Papp L. Plucinski, J. Minar, J. Braun, H. Ebert, C. M. Schneider and C. S. Fadley, *Band mapping in x-ray photoelectron spectroscopy: an experimental and theoretical study of W(110) with 1.25 keV excitation*, Phys. Rev. B **84**, 045433 (2011).
[5] C. S. Fadley, *X-ray photoelectron spectroscopy: Progress and perspectives,* Journal of Electron Spectroscopy and Related Phenomena **178–179,** 2 (2010).
[6] D. P. Woodruff, *Surface structural information from photo-electron diffraction*, J. Electron Spectrosc. Relat. Phenom. **178-179**, 186 (2010).
[7] J. Osterwalder *Photoelectron Spectroscopy and Diffraction,* in 'Handbook on Surface and Interface Science', ed. by K. Wandelt, Wiley-VCH, Weinheim, Vol. **1**, pp. 151-214 (2011).
[8] A. Winkelmann, C. S. Fadley and F. J. Garcia de Abajo, *High-energy photoelectron diffraction: model calculations and future possibilities*, New J. of Phys. **10,** 113002 (2008).
[9] G. Schönhense, K. Medjanik, S. Babenkov, D. Vasilyev, M. Ellguth, O. Fedchenko, S. Chernov, B. Schönhense and H.-J. Elmers, *Momentum-Transfer Model of Valence-Band Photoelectron Diffraction*, arXiv:1806.05871 (2018).

[10] S. V. Babenkov, Vi. Y. Aristov, O. V.Molodtsova, K. Winkler, L. Glaser, I. Shevchuk, F. Scholz, J. Seltmann, J. Viefhaus, *A new dynamic-XPS end-station for beamline P04 at PETRA III/DESY*, Nucl. Instr. and Methods in Phys. Research A **77**, 189 (2015).

[11] M. Hoesch, T. K. Kim, P. Dudin, H. Wang, S. Scott, P. Harris, S. Patel, M. Matthews, D. Hawkins, S. G. Alcock, T. Richter, J. J. Mudd, M. Basham, L. Pratt, P. Leicester, E. C. Longhi, A. Tamai, and F. Baumberger, *A facility for the analysis of the electronic structures of solids and their surfaces by synchrotron radiation photoelectron spectroscopy*, Review of Scientific Instruments **88**, 013106 (2017).

[12] S. Nemšák, S. Döring, C. Schlüter, M. Eschbach, E. Mlynczak, T.-L. Lee, L. Plucinski, J. Minar, J. Braun, H. Ebert, C. M. Schneider, C.S. Fadley, *Hard x-ray standing-wave angle-resolved photoemis-sion: element- and momentum-resolved band structure for a dilute magnetic semiconductor*, Nature Communications, in print: https://arxiv.org/abs/1801.06587.

[13] J. Tesch, F. Paschke, M. Fonin, M. Wietstruk, S. Böttcher, R.J. Koch, A. Bostwick, C. Jozwiak, E. Rotenberg, A. Makarova, B. Paulus, E. Voloshina and Y. Dedkov, *The graphene/n-Ge(110) interface: structure, doping, and electronic properties,* Nanoscale (2018), DOI: 10.1039/C8NR00053K.

[14] C. Schlueter, A. Gloskovskii, K. Ederer, S. Piec, M. Sing, R. Claessen, C. Wiemann, C.M. Schneider, K. Medjanik, G. Schönhense, P. Amann, A. Nilsson and W. Drube, *New HAXPES Applications at PETRA III*, Synchr. Radiation News **31**, 29 (2018).

[15] C. Tusche, A. Krasyuk and J. Kirschner, *Spin resolved bandstructure imaging with a high resolution momentum microscope*, Ultramicroscopy **159,** 520 (2015).

[16] G. Schönhense, A. Oelsner, O. Schmidt, G. H. Fecher, V. Mergel, O. Jagutzki, H. Schmidt-Böcking, *Time-Of-Flight Photoemission Electron Microscopy - A New Way To Chemical Surface Analysis*, Surf. Sci. **480**, 180-187 (2001).

[17] B. Schönhense, K. Medjanik, O. Fedchenko, S. Chernov, M. Ellguth, D. Vasilyev, A. Oelsner, J. Viefhaus, D. Kutnyakhov, W. Wurth, H. J. Elmers and G. Schönhense, *Multidimensional Photoemission Spectroscopy – the Space-Charge Limit*, New J. of Physics **20**, 033004 (2018).

[18] J. Viefhaus et al*., The Variable Polarization XUV Beam- line P04 at PETRA III: Optics, mechanics and their performance*, Nucl. Instrum. Methods **710**, 151-154 (2013).

[19] www.surface-concept.com, and private communication.

[20] B. Schönhense, G. Schönhense, K. Medjanik, D. Vasilyev, S. Babenkov, O. Fedchenko, S. Chernov, H. J. Elmers and K. Rossnagel, *Space charge in photoemission with pulsed sources: Strategies for correction and suppression*, to be submitted (2018).

[21] K. Medjanik et al., in preparation

[22] S. Chernov, K. Medjanik, D. Kutnyakhov, C. Tusche, S. A. Nepijko, A. Oelsner, J. Braun, J. Minár, S. Borek, H. Ebert, H. J. Elmers, J. Kirschner and G. Schönhense, *Anomalous d-like Surface Resonance on Mo(110) Analyzed by Time-of-Flight Momentum Microscopy*, Ultramicroscopy **159**, 463 (2015).

[23] C. Tusche, P. Goslawski, D. Kutnyakhov, M. Ellguth, K. Medjanik, H. J. Elmers, S. Chernov, R. Wallauer, D. Engel, A. Jankowiak, and G. Schönhense, *Multi-MHz Time-of-Flight Electronic Band-structure Imaging of Graphene on Ir(111)*, Appl. Phys. Lett. **108**, 261602 (2016).

[24] L. Plucinski, J. Minár, B. C. Sell, J. Braun, H. Ebert, C. M. Schneider and C. S. Fadley, *Band mapping in higher-energy x-ray photoemission: Phonon effects and comparison to one-step theory,* Phys. Rev. B **78**, 035108 (2008).

[25] Xu, S.-Y., N. Alidoust, I. Belopolski, Z. Yuan, G. Bian, T.-R. Chang, H. Zheng, V. N. Strocov, D. S. Sanchez, G. Chang, C. Zhang, D. Mou, Y. Wu, L. Huang, C.-C. Lee, S.-M. Huang, B. Wang, A. Bansil, H.-T. Jeng, T. Neupert, A. Kaminski, H. Lin, S. Jia and M. Zahid Hasan *Discovery of a Weyl fermion state with Fermi arcs in niobium arsenide*. Nat. Phys. **11**, 748 (2015).

[26] J. Osterwalder, T. Greber, S. Hüfner and L. Schlapbach, XPD *from a free-electron-metal valence band: Evidence for hole-state localization,* Phys. Rev. Lett. **64**, 2683 (1990).

[27] G. S. Herman, T. T. Tran, K. Higashiyama and C. S. Fadley, *Valence Photoelectron Diffraction and Direct Transition Effects*, Phys. Rev. Lett. **68**, 1204 (1992).

[28] J. Osterwalder, T. Greber, P. Aebi, R. Fasel and L. Schlapbach, *Final-state scattering in angle-resolved ultraviolet photoemission from copper*, Phys. Rev. B **53**, 10209 (1996).